S. L. Cherkas [1], V. L. Kalashnikov[2]

[1]*Institute for Nuclear Problems, Belarus State University, Minsk, Belarus*

[2]*Vienna University of Technology, Vienna, Austria*

# MATTER CREATION AND PRIMORDIAL CMB SPECTRUM IN THE INFLATIONLESS MILNE-LIKE COSMOLOGIES

We present a new insight into the interpretation of the primordial spectrum of scalar particles' density perturbations. On the assumption of spectrum universality, i.e., that the mean energy density and the typical value of inhomogeneity can be chosen arbitrarily in the framework of the model considered, the form of the spectrum becomes completely defined. It is close to the flat Harrison-Zeldovich's spectrum, but with the suppressed low-frequency modes.

## 1. Introduction

Lately, the Milne-like cosmologies attract a refreshing attention [1-8]. Although the original Milne's model [9] is open and empty and is not consistent with the radiation-dominated Big-Bang cosmological models, flat universes filled with some exotic matter with the overall equation of state $w = -1/3$ are actively scrutinized. Moreover, there is a deeper background for these cosmologies, relying on the residual vacuum fluctuations [10].

One of the attractive points for such a model is that it solves the horizon problem without inflation[i]. Recently, the most accurate cosmological data are based on the anisotropy of the cosmic microwave background (CMB). There is a well-known confrontation of the classical Milne-like cosmologies, excluding the radiation-dominated initial stage of the universe evolution, with experimental data, but such a confrontation is far from evidence in the framework of the quantum cosmology approach. In an inflation-less Milne-like cosmology, the issue is the description of the primordial spectrum of density perturbations. In the absence of inflation, we need a model for the primordial density perturbation spectrum, which differs from the inflation-based one but reproduces the CMB datum. Some steps have been done in this direction [7,11], but they refer to the methodology of the old inflationary paradigm and do not discuss a crucial issue of the matter origin in the universe. In the inflationary scenario, the matter appears as a result of the inflaton field decay, and if there is no inflation, then there is no inflaton field and no matter. However, the possible scenarios of universe evolution, primordial matter creation, and primordial inhomogeneity can differ from conventional ones, that is the matter of the issue of this work.

## 2. The wave packet description of the quantum field in the expanding universe

Quantum fields on the classical background were carefully considered earlier [12]. The main instruments for the description of the quantum field are operators of creation and annihilation used for quantization of the field oscillators. For the expanding universe, it seems, another formalism could be convenient. In fact, in a vicinity of singularity, there are no field oscillators because they do not begin to oscillate yet. On the other hand, it was demonstrated a finiteness of some quantities, namely momentums of the dynamical variables, despite an infinity of the dynamical variables itself in the singularity [13]. It is important that a state of quantum field can be described in terms of a wave packet over eigenfunctions of these momentums.

Let us begin from the conventional formalism for a scalar field $\phi(\eta,\mathbf{r})$ in the expanding universe described by the Lagrangian

$$L = \frac{1}{2}\int a^2 \left(\phi'^2 - (\nabla\phi)^2\right) d^3\mathbf{r}, \qquad (1)$$

where $a(\eta)$ is the scale factor.

The assumption, that the metric tensor of the universe corresponds to the interval

$$ds^2 = a^2(\eta)(d\eta^2 - d\mathbf{r}^2),$$

where $\eta$ is the conformal time, and the expansion of the scalar field in the Fourier series

$$\phi(\mathbf{r}) = \sum_\mathbf{k} \varphi_\mathbf{k} e^{i\mathbf{k}\mathbf{r}} \qquad (2)$$

allows rewriting the Lagrangian (1) as

$$L = \frac{a^2}{2}\sum_\mathbf{k} \varphi'_\mathbf{k}\varphi'_{-\mathbf{k}} - k^2 \varphi_\mathbf{k}\varphi_{-\mathbf{k}}. \qquad (3)$$

Corresponding momentums are $\pi_\mathbf{k} = \frac{\partial L}{\partial \varphi'_\mathbf{k}} = a^2 \varphi'_{-\mathbf{k}}$. In terms of momentums, the Hamiltonian $H = L - \sum_\mathbf{k}\varphi'_\mathbf{k}\frac{\partial L}{\partial \varphi'_\mathbf{k}}$ is

$$H = \frac{1}{2}\sum_\mathbf{k} \frac{\pi_\mathbf{k}\pi_{-\mathbf{k}}}{a^2} + a^2 k^2 \varphi_\mathbf{k}\varphi_{-\mathbf{k}}. \qquad (4)$$

Each $\mathbf{k}$-mode satisfies the equation of motion

$$\varphi''_\mathbf{k} + \frac{2a'}{a}\varphi'_\mathbf{k} + k^2 \varphi_\mathbf{k} = 0. \qquad (5)$$

Quantization in terms of creation and annihilation operators consists in postulating [12]

$$\hat{\varphi}_\mathbf{k} = \hat{a}^+_{-\mathbf{k}} u^*_k(\eta) + \hat{a}_\mathbf{k} u_k(\eta), \qquad (6)$$

where functions $u_k$ satisfy the equation of motion (5) and the relation

$$a^2(\eta)(u_k u'^*_k - u^*_k u'_k) = i. \qquad (7)$$

Let us prove that for a wide class of dependencies $a(\eta)$, such that the kinetic (first) term dominates the potential (second) term in Eq. (4), the momentums are finite quantities in the vicinity of the singularity. Momentum corresponding to the $\mathbf{k}$-mode can be written

$$\hat{\pi}_\mathbf{k} = a^2 \hat{\varphi}'_{-\mathbf{k}} = a^2(\eta)(\hat{a}_{-\mathbf{k}} u'_k(\eta) + \hat{a}^+_\mathbf{k} u'^*_k(\eta)). \qquad (8)$$

In the singularity neighborhood, the function $u_k(\eta)$ satisfies asymptotically the equation (5) without the last term. This equation can be represented in the form:

$$\frac{d}{d\eta}\left(a(\eta)^2 u_k^{'}(\eta)\right) = 0. \tag{9}$$

From the equations (8) and (9), one may conclude that the momentums $\hat{\pi}_{\mathbf{k}}$ are some constant operators in the singularity neighborhood asymptotically. The assertion about domination of the kinetic term in the singularity neighborhood is well-grounded, for instance, for dependencies $a(\eta) \sim \eta^n$. In particular, these dependencies include $a(\eta) \sim \eta$ (radiation background) and $a(\eta) \sim \eta^2$ (matter background). However, the assertion above is not valid for the Milne-like cosmology with $a(\eta) \sim \exp(H \eta)$. Nevertheless, we will argue in the next section that the Milne-like behavior begins not from the singularity but somewhat later.

Let us define the time-independent operators

$$\hat{P}_{\mathbf{k}} = \alpha_k \hat{a}_{-\mathbf{k}} + \alpha_k^* \hat{a}_{\mathbf{k}}^+, \tag{10}$$

where the complex constants are $\alpha_k = a^2(\eta) u_k^{'}(\eta)|_{\eta \to 0}$, and

$$\hat{X}_{\mathbf{k}} = b_k \exp(i\theta_k) \hat{a}_{\mathbf{k}} + b_k \exp(-i\theta_k) \hat{a}_{-\mathbf{k}}^+. \tag{11}$$

Then, imposing the commutation relations

$$[\hat{P}_{\mathbf{k}}, \hat{X}_{\mathbf{q}}] = -i\delta_{\mathbf{k},\mathbf{q}}. \tag{12}$$

allows finding the constants $b_k$. Taking into account that $[\hat{a}_{\mathbf{k}}, \hat{a}_{\mathbf{k}}^+] = 1$, we come to

$$b_k = -\frac{i}{\alpha_k e^{-i\theta_k} - \alpha_k^* e^{i\theta_k}}. \tag{13}$$

With the help of Eqs. (10), (11) and (13), we can express the creation and annihilation operators through $\hat{X}_{\mathbf{k}}$ and $\hat{P}_{\mathbf{k}}$:

$$\hat{a}_{\mathbf{k}} = \frac{\hat{P}_{\mathbf{k}}^+}{\alpha_k - e^{2i\theta_k}\alpha_k^*} - i\alpha_k^* \hat{X}_{\mathbf{k}}, \qquad \hat{a}_{-\mathbf{k}}^+ = \frac{e^{2i\theta_k}\hat{P}_{\mathbf{k}}^+}{\alpha_k^* e^{2i\theta_k} - \alpha_k} + i\alpha_k \hat{X}_{\mathbf{k}}. \tag{14}$$

Substitution of (14) into (6) leads to

$$\hat{\varphi}_{\mathbf{k}}(\eta) = \frac{\hat{P}_{\mathbf{k}}^+ \left(u_k(\eta) - e^{2i\theta_k} u_k^*(\eta)\right)}{\alpha_k - \alpha_k^* e^{2i\theta_k}} + i\hat{X}_{\mathbf{k}}(\alpha_k u_k^*(\eta) - \alpha_k^* u_k(\eta)). \tag{15}$$

The operators $\hat{P}_{\mathbf{k}}$ and $\hat{X}_{\mathbf{k}}$ is convenient to represent in the form $\hat{P}_{\mathbf{k}} = P_{\mathbf{k}}, \hat{X}_{\mathbf{k}} = i\frac{\partial}{\partial P_{\mathbf{k}}}$. Eq. (13) allows describing the quantum evolution of the fields using the wave packet $C(P_{\mathbf{q}})$ in the singularity neighborhood. For instance, the wave packet averaged value of $\hat{\varphi}_{\mathbf{k}}(\eta)$ takes the form

$$<\psi|\hat{\varphi}_{\mathbf{k}}(\eta)|\psi> = \frac{e^{-i\theta_k} u_k(\eta) - e^{i\theta_k} u_k^*(\eta)}{e^{-i\theta_k}\alpha_k - e^{i\theta_k}\alpha_k^*} \int (C(P_{\mathbf{q}}))^* P_{\mathbf{k}}^* C(P_{\mathbf{q}}) \mathbf{D} P_{\mathbf{q}} \mathbf{D} P_{\mathbf{q}}^*$$
$$-\left(\alpha_k u_k^*(\eta) - u_k(\eta)\alpha_k^*\right) \int (C(P_{\mathbf{q}}))^* \frac{\partial}{\partial P_{\mathbf{k}}} C(P_{\mathbf{q}}) \mathbf{D} P_{\mathbf{q}} \mathbf{D} P_{\mathbf{q}}^*, \tag{16}$$

where the integration implies $\mathbf{D} P_{\mathbf{k}} = dP_0 dP_{\mathbf{k}_1} dP_{\mathbf{k}_1}^* dP_{\mathbf{k}_2} dP_{\mathbf{k}_2}^* \ldots$ The integral over $dz dz^* \equiv \frac{\rho d\rho d\phi}{2\pi i}$, $z = \rho e^{i\phi}$ is understood in the holomorphic representation [14].

Let us consider the transformation of the wave packet $C(P_\mathbf{k}) \to C(P_\mathbf{k})\exp\left(-i\sum_\mathbf{q} g_q P_\mathbf{q}^* P_\mathbf{q}\right)$ in Eq. (16), where $g_k$ are some real constants. Under summation on $\mathbf{q}$ one should take into account that $P_{-\mathbf{q}} = P_\mathbf{q}^*$. As can see, the transformation is equivalent to changing the phase $\theta_k$ so that

$$\theta_k \to \arctan\left(\frac{\tan\theta_k \left(2 + ig_k\left(\alpha_k^2 - \alpha_k^{*2}\right)\right) - g_k(\alpha_k - \alpha_k^*)^2}{2 - g_k\left((\alpha_k + \alpha_k^*)^2 \tan\theta_k + i\left(\alpha_k^2 - \alpha_k^{*2}\right)\right)}\right). \tag{17}$$

These phases $\theta_k$ could be considered twofold: (i) mathematically, they are the phases of the basis functions $u_k(\eta)$, but (ii) physically, they are the property of the quantum state in the chosen basis $u_k$, because they are equivalent to the constants $g_k$. Further, we will consider a Gaussian wave packet $C(P_\mathbf{k}) = \exp\left(-\sum_\mathbf{q} \Delta_\mathbf{q} P_\mathbf{q}^* P_\mathbf{q}\right)$ with the real constant $\Delta_\mathbf{k}$ as a width, and a phase $\theta_k$ of $u_k(\eta)$. Let's emphasize, that the matter is *encoded in the singularity* that it is not created from the vacuum. Even for a massive particle, the amount of matter created from the vacuum is not sufficient to explain the universe contents in the Milne-like cosmology [15] for the observed conformal Hubble constant $\mathbf{H}$.

### 3. Model of the universe expansion

It is evident that the Milne-like behavior is hardly extendable to the vicinity of singularity. Let us consider a heuristic model to describe the way in which the universe could come to the Milne-like expansion. Consider Hamiltonian in which the universe metric is uniform, but the non-uniform scalar field presents [10,16]:

$$H = -\frac{1}{2}M_p^2 a'^2 + \frac{1}{2}\sum_\mathbf{k} \frac{\pi_\mathbf{k} \pi_\mathbf{k}^+}{a^2} + a^2 k^2 \phi_\mathbf{k} \phi_\mathbf{k}^+. \tag{18}$$

Initially, the last term corresponding to the potential energy of the field oscillators does not play a role. Consequently, $a(\eta) \sim \sqrt{\eta}$ under the small conformal time, because $a'^2 \sim \dfrac{\pi_\mathbf{k} \pi_\mathbf{k}^*}{a^2}$ and $\pi_\mathbf{k} \sim const$. When the field oscillators begin to oscillate, the expansion changes from $a(\eta) \sim \sqrt{\eta}$ to $a(\eta) \sim \exp(\mathbf{H}\eta)$, i.e., $a(t) \sim \mathbf{H} t$ in cosmic time [10]. The late-time rate of the universe expansion also differs from the Milne-like, because the universe accelerates due to residual vacuum fluctuations [10]. Here, for simplicity, we will not take this acceleration into account.

Due to the UV cut-off of the comoving momentums, high energy oscillators at the Planck frequencies give the main contribution to the universe expansion [10,16]. Thus, the change of the expansion rate occurs near the Planck time and can be described by the model

$$a(\eta) = \begin{cases} \gamma\sqrt{\eta} + \beta\eta, & 0 < \eta < \eta_1, \\ B\exp(\mathbf{H}\eta), & \eta > \eta_1, \end{cases} \tag{19}$$

where $\eta_1$ is of the order of the Planck time. For the smooth splicing of the function and their derivatives, the linear term $\beta\eta$ was introduced in Eq. (19). Values of the coefficients equal

$$\gamma = \frac{2Be^{\mathbf{H}\eta_1}(1-\mathbf{H}\eta_1)}{\sqrt{\eta_1}}, \qquad \beta = -\frac{Be^{\mathbf{H}\eta_1}(1-2\mathbf{H}\eta_1)}{\eta_1}. \tag{20}$$

The Milne-like models seem to be attractive since they could solve the horizon problem without inflation, but this prospect puts the constraint on the coefficient $B$ in Eqs. (19), (20). One of the formulations of the horizon problem is that the last scattering surface consists of some causally disconnected regions. Let $\eta_0$ will be today number of conformal time corresponding to the present-day scale factor $a_0$. The horizon is the present observable part of the universe which could be reached by light:

$$R(t_0) = a(t_0)\int_0^{t_0}\frac{dt'}{a(t')}. \tag{21}$$

In comoving coordinates $R(t_0)/a(t_0)$, the horizon is simply $\int_0^{\eta_0}d\eta = \eta_0$, i.e., the conformal time from the beginning of the universe. Thus, $R(t_0) = a_0\eta_0$. The size of the region corresponding to the present horizon at the time of last scattering is $R(t_0)\frac{a_{LS}}{a_0}$. This size should be an order of or less than the horizon at the last scattering time $R(t_{LS})$:

$$R(t_0)\frac{a_{LS}}{a_0} \leq R(t_{LS}) = a_{LS}\eta_{LS}, \tag{22}$$

otherwise it will consist of some causally disconnected regions. Substitution of $\eta_0 = \frac{1}{\mathbf{H}}\ln\frac{a_0}{B}$ and $\eta_{LS} = \frac{1}{\mathbf{H}}\ln\frac{a_{LS}}{B}$ to Eq. (22) gives

$$\ln\frac{a_0}{B} \approx \ln\frac{a_{LS}}{B}, \tag{23}$$

where the sign $\leq$ is changed by $\approx$ because the first one is impossible in principle. The redshift of the last scattering surface is $z_{LS} \approx 1100$, i.e. $a_0/a_{LS} \approx 10^{-2}$, thus $B$ must be sufficiently small. For instance, taking $a_0 \equiv 1$ and $B = 10^{-30}$ one has $\ln 10^{30} \approx \ln 10^{28}$. The problem of the horizon is solved in linear cosmologies by the minimal way: although the region at last scattering surface corresponding to the present horizon is less than the horizon at last scattering, they equal approximately if the constant $B$ is sufficiently small.

Solutions of Eqs. (5) (although this equation is for $\varphi_\mathbf{k}$, $u_k$ obeys it as well) in the different regions can be written as

$$u_k(\eta) = \begin{cases} \dfrac{\sqrt{\pi}}{2\gamma}\left(1 + \dfrac{2i}{\pi}\left(\dfrac{2\gamma}{\gamma+\beta\sqrt{\eta}} - 2\ln\left(\gamma+\beta\sqrt{\eta}\right) + \ln(\eta)\right)\right), & 0 < \eta < \eta_1, \\[2mm] \dfrac{e^{-\mathbf{H}\eta+i\vartheta-i\eta\sqrt{k^2-\mathbf{H}^2}}}{\sqrt{2B}\sqrt[4]{k^2-\mathbf{H}^2}}, & \eta > \eta_1, \end{cases} \tag{24}$$

where the phase $\vartheta = -\arctan\frac{\sqrt{k^2-\mathbf{H}^2}}{\mathbf{H}} + \eta_1\sqrt{k^2-\mathbf{H}^2}$. Since $\eta_1$ is very small versus $1/k$, an approximate non-oscillating solution is taken in the expression (24) at $0<\eta<\eta_1$. Besides, there could be an overall phase of the functions $u_k$ which influences the matter production through the quantities $\alpha_k$.

### 4. Mean energy density

Let us calculate the mean energy density of created particles defined as

$$\bar{\rho} = <C[P]|\hat{\rho}|C[P]> - <0|\hat{\rho}|0>, \tag{25}$$

where an average vacuum density is subtracted and

$$\hat{\rho} = \frac{1}{V}\int_V\left(\frac{\hat{\phi}'^2}{2a^2} + \frac{(\nabla\hat{\phi})^2}{2a^2}\right)d^3\mathbf{r} = \frac{1}{2a^2}\sum_{\mathbf{k}}\hat{\phi}'_{\mathbf{k}}\hat{\phi}'_{-\mathbf{k}} + k^2\hat{\phi}_{\mathbf{k}}\hat{\phi}_{-\mathbf{k}}, \tag{26}$$

where $V$ is the normalization volume. Substitution of the expression (15) into (26), taking into account that $\hat{\phi}_{-\mathbf{k}} = \hat{\phi}^+_{\mathbf{k}}$, and performing the integration over $\mathbf{D}P_\mathbf{q}\mathbf{D}P^*_\mathbf{q}$ results in

$$\bar{\rho} = \frac{1}{a^2}\sum_{|\mathbf{k}|>\mathbf{H}}\frac{1}{4\Delta_k(\alpha_k-\alpha^*_k)^2}\Big(k^2 u_k(\eta)^2 - 2k^2 u_k(\eta)u^*_k(\eta) + k^2 u^*_k(\eta)^2 + u'_k(\eta)^2 - 2u'_k(\eta)u'^*_k(\eta)$$
$$+ u'^*_k(\eta)^2\Big) - \frac{\Delta_k}{4}\Big(\alpha^{*2}_k k^2 u_k(\eta)^2 - 2\alpha_k\alpha^*_k k^2 u_k(\eta)u^*_k(\eta) + \alpha^2_k k^2 u^*_k(\eta)^2 \tag{27}$$
$$+ \big(\alpha^*_k u'_k(\eta) - \alpha_k u'^*_k(\eta)\big)^2\Big) - \frac{1}{2}\Big(u'^*_k(\eta)u'_k(\eta) + k^2 u^*_k(\eta)u_k(\eta)\Big),$$

where the last term corresponds to the vacuum average, which we subtract accordingly Eq. (25). It should be noted, that the summation over $\mathbf{k}$ in Eq. (27) is limited by the value of the conformal Hubble constant. That is ad hoc assumption because we discuss the creation of the particles on the vacuum background. The modes with $|\mathbf{k}|<\mathbf{H}$ do not oscillate and do not correspond to real particles.

Let us consider not only quantum average but also time average for the concrete form of the functions $u_k(\eta)$ given by (24). That causes a further simplification because it removes oscillating terms. As a result, we come to

$$\bar{\rho} = \frac{1}{2a^4}\sum_{|\mathbf{k}|>\mathbf{H}}\frac{k^2}{\sqrt{k^2-\mathbf{H}^2}}\left(\alpha_k\alpha^*_k\Delta_k - \frac{1}{(\alpha_k-\alpha^*_k)^2\Delta_k} - 1\right). \tag{28}$$

Minimization of the energy density by choosing the corresponding $\Delta_k$, i.e., $\Delta_k = |\alpha_k(\alpha^*_k-\alpha_k)|^{-1}$, gives

$$\bar{\rho} = \frac{1}{2a^4}\sum_{|\mathbf{k}|>\mathbf{H}}\frac{k^2}{\sqrt{k^2-\mathbf{H}^2}}\left(2\alpha^*_k\alpha_k\sqrt{-(\alpha^*_k-\alpha_k)^2}-1\right). \tag{29}$$

Quantities $\alpha_k$ characterize phases of the functions $u_k(\eta)$ near the singularity. It is convenient to represent them in the form $\alpha_k = r_k\exp(i(\pi/2+\theta_k))$, where $r_k$ and $\theta_k$ are reals.

As a result, we come to the final formula for the density of particles created from the wave packet set in the singularity

$$\bar{\rho} = \frac{1}{2a^4} \sum_{|\mathbf{k}|>\mathbf{H}} \frac{k^2}{\sqrt{k^2-\mathbf{H}^2}}(\sec\theta_k - 1) = \frac{1}{4\pi^2 a^4} \int_{\mathbf{H}}^{\infty} \frac{k^4}{\sqrt{k^2-\mathbf{H}^2}}(\sec\theta_k - 1) dk, \qquad (30)$$

where the integration is changed by the summation $\sum_{\mathbf{k}} \to \int \frac{d^3\mathbf{k}}{(2\pi)^3}$.

Let us consider the model which allows obtaining arbitrary mean energy density taking

$$\sec\theta_k = 1 + (\mu k)^n, \qquad (31)$$

where $\mu$ and $n$ are some constants. As seen from (31), the energy density is proportional to $\mu^n$ and, thus, can be set arbitrary. Besides, we will impose the UV cut-off on the momentums of the order of the Planck mass, if some divergent integrals appear [10,16]. It was shown, that this cut-off is necessary to obtain a true value of the universe acceleration due to residual vacuum fluctuations [16].

### 5. The primordial spectrum of the energy density inhomogeneity

Let us calculate the spectrum of the energy density

$$\hat{\rho}(\mathbf{r}) = \frac{1}{2a^2}\left(\hat{\phi}'(\mathbf{r})^2 + (\nabla\hat{\phi}(\mathbf{r}))^2\right). \qquad (32)$$

As we consider the Gaussian wave packets $C$, the energy density in the different space points takes the form

$$<C|\hat{\rho}(\mathbf{r})\hat{\rho}(\mathbf{r}')|C> = \zeta(|\mathbf{r}-\mathbf{r}'|), \qquad (33)$$

where

$$\zeta(r) = \sum_{\mathbf{k}} \sigma_k^2 \exp(i\mathbf{k}\mathbf{r}), \qquad \sigma_k^2 = <C|\hat{\rho}_{\mathbf{k}}^+\hat{\rho}_{\mathbf{k}}|C> - <0|\hat{\rho}_{\mathbf{k}}^+\hat{\rho}_{\mathbf{k}}|0>. \qquad (34)$$

Let's emphasize that a normal ordering is not used in (34) because the summation covers all $\mathbf{k}$-domain and $\rho_{-\mathbf{k}} = \rho_{\mathbf{k}}^+$. In Eq. (34) we again subtract the vacuum average ad hoc, because of only the spectrum of the created particles on the vacuum background is of interest.

The Fourier components of energy density are expressed as

$$\hat{\rho}_{\mathbf{k}} = \frac{1}{V}\int_V \hat{\rho}(\mathbf{r})\exp(-i\mathbf{r}\mathbf{k})d^3\mathbf{r} = \frac{1}{2a^2}\sum_{\mathbf{q}}\hat{\phi}_{\mathbf{q}}^{+'}\hat{\phi}_{\mathbf{q}-\mathbf{k}}' + (\mathbf{q}-\mathbf{k})\mathbf{q}\hat{\phi}_{\mathbf{q}}^+\hat{\phi}_{\mathbf{q}-\mathbf{k}},$$

$$\hat{\rho}_{\mathbf{k}}^+ = \frac{1}{2a^2}\sum_{\mathbf{q}}\hat{\phi}_{\mathbf{q}-\mathbf{k}}^{+'}\hat{\phi}_{\mathbf{q}}' + (\mathbf{q}-\mathbf{k})\mathbf{q}\hat{\phi}_{\mathbf{q}-\mathbf{k}}^+\hat{\phi}_{\mathbf{q}}. \qquad (35)$$

We remind the steps of the previous section: (i) integration over $\mathbf{D}P_{\mathbf{q}}\mathbf{D}P_{\mathbf{q}}^*$, using $\Delta_k$ corresponding to the minimal energy density and (ii) removing the oscillating terms. Here the calculations are the same, but more complicated, and can be done using the Mathematica software. As a result, with the functions $u_k(\eta)$ given by (34), we come to

$$\sigma_k^2 = <C|\hat{\rho}_{\mathbf{k}}\hat{\rho}_{\mathbf{k}}^+|C> - <0|\hat{\rho}_{\mathbf{k}}\hat{\rho}_{\mathbf{k}}^+|0> =$$

$$\sum_{\substack{|\mathbf{q}|>\mathbf{H}, \\ |\mathbf{q}-\mathbf{k}|>\mathbf{H}}} \frac{\left(\mathbf{q}(\mathbf{q}-\mathbf{k})\left(2\mathbf{H}^2 + \mathbf{q}(\mathbf{q}-\mathbf{k})\right) + q^2(\mathbf{q}-\mathbf{k})^2\right)(\sec\theta_q \sec\theta_{|\mathbf{q}-\mathbf{k}|} - 1)}{16a^8\sqrt{(q^2-\mathbf{H}^2)((\mathbf{q}-\mathbf{k})^2-\mathbf{H}^2)}}, \qquad (36)$$

where a summation over **q** is again confined by considering only real particles. It is known, that the inhomogeneity of CMB is relatively small. That restricts the value of the relative inhomogeneity given by the dimensionless quantity $k^3\sigma_k^2/\bar{\rho}^2$. Let us again to take the dependence (31). According to (30), $\bar{\rho}$ is proportional to the constant $\mu^n$. At $n=4$ the integral in Eq. (30) diverges logarithmically. In the general case, one has:

$$\bar{\rho} \sim \begin{cases} \dfrac{1}{a^4}\mathbf{H}^4\left(\dfrac{\mu}{\mathbf{H}}\right)^n\left(\dfrac{k_{max}}{\mathbf{H}}\right)^{4-n}, & n<4, \\[2mm] \dfrac{1}{a^4}\mathbf{H}^4\left(\dfrac{\mu}{\mathbf{H}}\right)^4 \ln\left(\dfrac{k_{max}}{\mathbf{H}}\right), & n=4, \\[2mm] \dfrac{1}{a^4}\mathbf{H}^4\left(\dfrac{\mu}{\mathbf{H}}\right)^n, & n>4, \end{cases} \qquad (37)$$

where $k_{max}$ is the UV cut-off of the order of the Planck mass. The analogous estimation for Eq. (36), when the summation $\sum_\mathbf{q}$ is replaced by the integration $\int \dfrac{d^3\mathbf{q}}{(2\pi)^3}$, leads to

$$k^3\sigma_k^2\big|_{k\sim\mathbf{H}} \sim \begin{cases} \dfrac{\mathbf{H}^8}{a^8}\left(c_1\left(\dfrac{k_{max}}{\mathbf{H}}\right)^{5-n}\left(\dfrac{\mu}{\mathbf{H}}\right)^n + c_2\left(\dfrac{\mu}{\mathbf{H}}\right)^{2n}\right), & 2<n<5, \\[2mm] \dfrac{\mathbf{H}^8}{a^8}\left(c_1\left(\dfrac{\mu}{\mathbf{H}}\right)^n + c_2\left(\dfrac{\mu}{\mathbf{H}}\right)^{2n}\right), & n>5. \end{cases} \qquad (38)$$

The present temperature of CMB is $T_0 = 2.73\,K = 2.35\times10^{-4}\,eV$, the UV cut-off is of the order of the Planck mass $k_{max} \sim M_p = \sqrt{\dfrac{3}{4\pi G}} = 6\times10^{18}\,GeV$, and the Hubble constant is $\mathbf{H} \sim 2.1\times10^{-33}\,eV$. Energy density corresponding to the CMB temperature is

$$\bar{\rho} \sim \dfrac{1}{a^4} T_0^4. \qquad (39)$$

Using Eqs. (37), (39), we can rewrite Eq. (38) regarding $T_0$

$$\frac{k^3 \sigma_k^2}{\bar{\rho}^2}\bigg|_{k\sim\mathbf{H}} \sim \begin{cases} c_1 \dfrac{k_{max}}{\mathbf{H}} \left(\dfrac{T_0}{\mathbf{H}}\right)^{-4} + c_2 \left(\dfrac{k_{max}}{\mathbf{H}}\right)^{2n-8}, & 2 < n < 4, \\[2mm] c_1 \dfrac{k_{max}}{\mathbf{H}} \left(\ln\left(\dfrac{k_{max}}{\mathbf{H}}\right)\right)^{-1} \left(\dfrac{T_0}{\mathbf{H}}\right)^{-4} + c_2 \left(\ln\left(\dfrac{k_{max}}{\mathbf{H}}\right)\right)^{-2}, & n = 4, \\[2mm] c_1 \left(\dfrac{k_{max}}{\mathbf{H}}\right)^{5-n} \left(\dfrac{T_0}{\mathbf{H}}\right)^{-4} + c_2, & 4 < n < 5, \\[2mm] c_1 \left(\dfrac{T_0}{\mathbf{H}}\right)^{-4} + c_2, & n > 5, \end{cases} \qquad (40)$$

where $c_1$ and $c_2$ are some constants of the order of unity. The first observation is that the $c_1$-term is always suppressed by the multiplier $(T_0/\mathbf{H})^{-4}$ and is negligible, despite the presence of the large multiplier $k_{max}/\mathbf{H}$. The second observation is that the low relative inhomogeneity could be obtained if $n$ is close to 4. In particular, $\dfrac{k^3 \sigma_k^2}{\bar{\rho}^2} \sim (\ln(k_{max}/\mathbf{H}))^{-2} \sim 2 \times 10^{-4}$ for $n = 4$.

The next step is the calculation of a spectrum form. For low inhomogeneity, it is completely fixed by (31), because $n$ should be very close to 4 and we come to the integral

$$\sigma_k^2 = \frac{\mu^{2n}}{16 a^8} \int_{|\mathbf{q}|>\mathbf{H},\,|\mathbf{q}-\mathbf{k}|>\mathbf{H}} \frac{\left(\mathbf{q}(\mathbf{q}-\mathbf{k})\left(2\mathbf{H}^2 + \mathbf{q}(\mathbf{q}-\mathbf{k})\right) + q^2(\mathbf{q}-\mathbf{k})^2\right)}{q^n (\mathbf{q}-\mathbf{k})^n \sqrt{(q^2-\mathbf{H}^2)((\mathbf{q}-\mathbf{k})^2 - \mathbf{H}^2)}} \frac{d^3\mathbf{q}}{(2\pi)^3}, \qquad (41)$$

which can be calculated by the Monte-Carlo method. For $n \approx 4$ it is convergent at the upper limit and contains the only parameter $\mathbf{H}$. Spectrum $\mathcal{P}(k) = \dfrac{k^3 \sigma_k^2}{\bar{\rho}^2}$ is shown in figure 1.

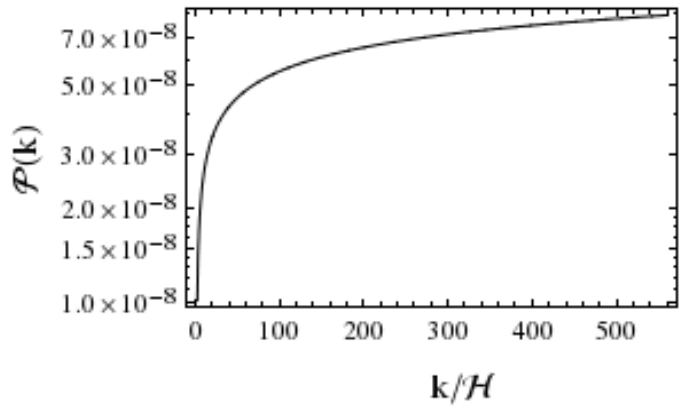

Fig. 1. The primordial spectrum of the relative density inhomogeneity $\mathcal{P}(k) = \dfrac{k^3 \sigma_k^2}{\bar{\rho}^2}$ for the parameter $n = 3.95$ in Eq. (31).

The spectrum is relatively flat but with the suppression of the low comoving momentums $k$. It is interesting that it is possible to obtain even more flat spectrum shown in figure 2 for $n = 4.11$, but the value of the relative inhomogeneity is large. For a small inhomogeneity, the parameter $n$ should be a bit less than four.

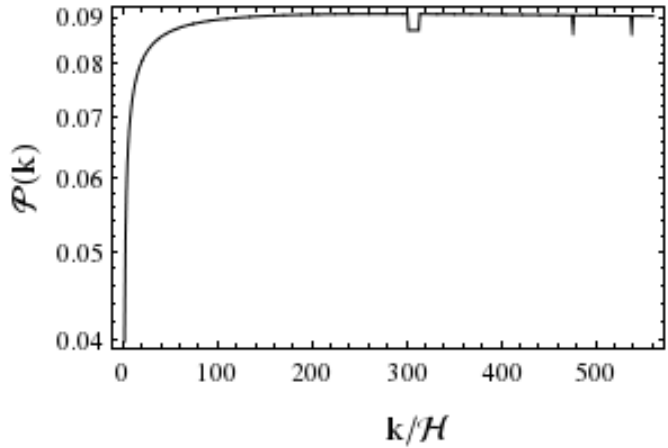

Fig. 2. The primordial spectrum of the relative density inhomogeneities $\mathbf{P}(k) = \dfrac{k^3 \sigma_k^2}{\bar{\rho}^2}$ for the parameter $n = 4.11$ in Eq. (31).

### 6. Conclusions

Studies of the Milne-like models seem attractive because they do not demand inflation. Moreover, all scales of interest always remain within the horizon during the evolution of the universe. Let's remind that in the standard model of the radiation dominant universe with the inflation stage, a mode crosses horizon at the stage of inflation and then returns at the radiation or matter domination stage.

It is shown that there exists the possibility to put any amount of matter into the cosmological singularity. If one demands that the relative value of inhomogeneity is small, the primordial spectrum has the fixed form, with the suppression of the large-scale modes. Because this spectrum corresponds to the scalar particles, the most evident candidate is the Higgs particles, which has been discovered recently. These particles decay into the photons and transfer the inhomogeneity to them. That is, the spectrum of scalar particles gives the initial conditions for the evolution of the photon modes. The initial conditions should be set at the moment of Higgs decay. The next step is to investigate the evolution of the photon spectrum and the massive matter spectrum in the Milne-like cosmology, but it is beyond the scope of the present work. Besides, there could be an analogous spectrum of the gravitational waves, which are equivalent to the massless scalar fields [16] but do not undergo any change except for the cooling due to universe expansion.

### Acknowledgements

The authors are grateful to the organizers of the First ICRANet-Minsk workshop on high energy astrophysics in a frame of BelINP-2017 Symposium.

---

[i] The Milne-like cosmological models are, in principle, are compatible with the inflation paradigm but without tacking into account the quantum effects [17].